\documentclass{article}
\usepackage{amssymb}

\usepackage{graphicx}
\usepackage{amsmath}

\begin{document}

\title{Entanglement bounds of tripartite squeezed thermal states }
\author{Xiao-yu Chen \\
%EndAName
School of Science, China Institute of Metrology,\\
310018,Hangzhou, China}
\date{}
\maketitle

\begin{abstract}
I propose the multi-mode squeezed thermal state based on the
multi-mode pure entangled state. The correlation matrix of the
state is characterized by two parameters. I then analysis the
separable condition for this state, and calculating the relative
entropy of the state with respect to the same kind of fully
separable state in order to provide an upper bound of the relative
entropy of entanglement. The bound is compared with the other
bounds which were obtained with reduced state.
\end{abstract}

\section{Introduction}

The quantification of entanglement is one of the most important
problem in quantum information theory. The three most promising
entanglement measures are the entanglement of formation, the
entanglement of distillation \cite {Bennett} and the relative
entropy of entanglement \cite{Vedral1}. In the bipartite system,
the meaning of both entanglement of formation and the entanglement
of distillation is quite clear. However, these two measures do not
have an entirely straightforward meaning when one considers
multi-partite entanglement \cite{Vedral2}. On the other hand, the
relative entropy of entanglement can easily be generalized to
multi-partite system. Recently, towards possible applications in
quantum communication, both theoretical and experimental
investigations increasingly focus on quantum states with a
continuous spectrum defined in an infinite dimensional Hilbert
space. These states can be relatively easily generated using
squeezed light and beam splitters. Besides the research on the
entanglement of bipartite continuous variable system, the non-
locality of the multipartite entangled continuous-variable
Greenberger-Horne-Zeilinger (CVGHZ) states was proved
\cite{Loock}. Hence the quantification of the entanglement of
CVGHZ states comes into our sight. Now let us proceed to derive a
simple CVGHZ mixed state and give out its upper bound of the
relative entropy of entanglement.

\section{Multi-mode squeezed thermal state}

The continuous variable analogues to the multipartite entangled
Greenberger-Horne-Zeilinger states was proposed\cite{Loock}. The
Wigner function of the pure entangled m-mode state is (with $\hbar
=1$)

\begin{eqnarray}
W(\mathbf{x,p)} &=&\left( \frac{1}{\pi }\right) ^{m}\exp \{-e^{-2r}\left[
\frac{1}{m}\left( \sum_{i=1}^{m}x_{i}\right) ^{2}+\frac{1}{2m}%
\sum_{i,j}^{m}\left( p_{i}-p_{j}\right) ^{2}\right]   \notag \\
&&-e^{2r}\left[ \frac{1}{m}\left( \sum_{i=1}^{m}p_{i}\right) ^{2}+\frac{1}{2m%
}\sum_{i,j}^{m}\left( x_{i}-x_{j}\right) ^{2}\right] \}
\end{eqnarray}
The correlation matrix (CM) of this state is $\alpha =\alpha
_{x}\oplus \alpha _{p}.$ The $m\times m$ matrices $\alpha _{x}$
and $\ \alpha _{p}$ are in a special kind of form. If we define a
$m\times m$ matrix $A(x,y)$ with its all diagonal elements being
equal to $\frac{1}{m}(x+\left( m-1\right) y), $ all off-diagonal
elements being equal to $\frac{1}{m}(x-y),$ then $\alpha
_{x}=\frac{1}{2}A\left( e^{2r},e^{-2r}\right) ,$ $\alpha _{p}=\frac{1}{2}%
A\left( e^{-2r},e^{2r}\right) $. It is easy to check that
\begin{equation}
A(x,y)A\left( u,v\right) =A(xu,yv).  \label{wave1}
\end{equation}
So that \ $\alpha _{x}\alpha _{p}=\alpha _{p}\alpha _{x}=\frac{1}{4}I_{m}$.

Now we introduce the m-mode squeezed thermal state (mST) based on the pure
entangled m-mode state. The CM of mST state is $\alpha =\alpha _x\oplus
\alpha _p$ with
\begin{equation}
\alpha _x=\left( N+\frac 12\right) A\left( e^{2r},e^{-2r}\right) ,\text{ \ \
}\alpha _p=\left( N+\frac 12\right) A\left( e^{-2r},e^{2r}\right) .
\label{wave2}
\end{equation}
where $N$ is the average photon number of thermal noise of every mode. Then
we have \
\begin{equation}
\alpha _x\alpha _p=\alpha _p\alpha _x=\left( N+\frac 12\right) ^2I_m.
\end{equation}
In the calculation of relative entropy between these Gaussian
states, We need the so called $M$ matrix which is the matrix in
the exponential expression of density operator of Gaussian
state\cite{Chen}\cite{Scheel}. It is related to the CM by a
symplectic transformation $S$. The CM can be diagonalized by
symplectic transformation such that $\alpha =S(\frac 12\coth
\frac 12\widetilde{M})S^T=S\widetilde{\alpha }S^T$ and $S^TMS=$ $\widetilde{M%
}$ (with $\widetilde{\alpha },\widetilde{M}$ being diagonal). The procedure
of obtaining the $M$ matrix is to explore the eigenfunctions of matrix $%
\alpha _x\alpha _p$ of the state\cite{Chen}. But an unit matrix
form of the matrix $\alpha _x\alpha _p$ like in Eq. (\ref{wave2})
has not definite eigenfunctions. Here the $M$ matrix is
constructed directly by making use of Eq. (\ref{wave1}).Clearly
the mST state we introduced is symmetric to all $m$ modes. So that
$\widetilde{\alpha }$ will be in the form of

\begin{equation}
\widetilde{\alpha }=\left( N+\frac{1}{2}\right) \left( I_{m}\oplus
I_{m}\right) ,
\end{equation}
and $\alpha =\left( N+\frac{1}{2}\right) SS^{T}.$ From Eq. (\ref{wave1}), we
can write Eq.(\ref{wave2}) as
\begin{equation}
\alpha _{x}=\left( N+\frac{1}{2}\right) A\left( e^{r},e^{-r}\right) A\left(
e^{r},e^{-r}\right) ,
\end{equation}
\ and
\begin{equation}
\alpha _{p}=\left( N+\frac{1}{2}\right) A\left( e^{-r},e^{r}\right) A\left(
e^{-r},e^{r}\right) .
\end{equation}
Hence

\begin{equation}
S=A\left( e^{r},e^{-r}\right) \oplus A\left( e^{-r},e^{r}\right) .
\end{equation}
The $M$ matrix will be

\begin{equation}
M=A\left( e^{-r},e^r\right) \log \frac 1vI_mA\left( e^{-r},e^r\right) \oplus
A\left( e^r,e^{-r}\right) \log \frac 1vI_mA\left( e^r,e^{-r}\right) ,
\label{wave3}
\end{equation}
where $v=\frac N{N+1}.$ The mST state we introduced is
characterized by two parameters $r$ and $N.$ When one of them is
$0,$the other one has a clear physical meaning: quantum
correlation or noise. And when both of them are not $0$, we will
display with three mode state that it is the competition of the
two parameters which will determine the state is authentically
multipartite entangled or not.

\section{Separability of the $1\times 1\times 1$ squeezed thermal state}

The separability problem of the three mode gaussian state was
perfectly solved\cite{Giedke}. I in this paper apply it in order
to derive a simple criterion in the special case of 3ST
state($1\times 1\times 1$ squeezed thermal state). The three mode
gaussian states were classified as 5 different entangled classes.
But 3ST states can be classified as 3 different entangled classes:
fully inseparable states, biseparable states, fully
separable states. Following the notation of Ref. \cite{Giedke}, the CM $%
\alpha $ now is replaced with $\gamma $, where $\gamma =2\alpha $, and
denote by $\Lambda _x=\Lambda \oplus I$ the partial transposition in $x^{,}$%
s system only, $x=A,B,C$ is one of the three modes. Denoting the partially
transposed CM by $\widetilde{\gamma }_x=\Lambda _x\gamma \Lambda _x,$ and
\begin{equation}
J=\left[
\begin{array}{cc}
0 & -I_m \\
I_m & 0
\end{array}
\right] .
\end{equation}
The criterion for fully inseparable state is

\begin{equation}
\widetilde{\gamma }_x\ngeq iJ,\text{ for all }x=A,B,C.
\end{equation}
Because of the symmetry of the 3ST state, the criterion can be
simplified to for example $\widetilde{\gamma }_A\ngeq iJ$ . This
will lead to the condition that

\begin{equation}
\cosh ^22r<\frac 9{32}\left( 2N+1+\frac 1{2N+1}\right) ^2-\frac 18.
\label{wave4}
\end{equation}
While for $\widetilde{\gamma }_x\geq iJ,($ $x=A,B,C).$ The state
which we will call it PPT trimode state can be biseparable or
fully separable. The criterion to distinguish the biseparable and
fully separable states is \cite {Giedke} as follow. The CM $\gamma
$ of PPT trimode state can be written of as

\begin{equation}
\gamma =\left(
\begin{array}{ll}
A & C \\
C^T & B
\end{array}
\right) ,
\end{equation}
where $A$ is a $2\times 2$\ CM for the first mode, whereas $B$ is a $4\times
4$ CM for the other two modes. Define the matrices $K$ and $\widetilde{K}$ as

\begin{equation}
K\equiv A-C\frac 1{B-iJ}C^T,\text{ \qquad }\widetilde{K}\equiv A-C\frac 1{B-i%
\widetilde{J}}C^T,
\end{equation}
where $\widetilde{J}=J\oplus \left( -J\right) $ is the partially transposed $%
J$ for two modes.

Then the condition of the PPT trimode state to be fully separable is that if
and only if there exists a point $(y,z)\in R^2$ fulfilling the following
inequality:

\begin{eqnarray}
\min \{trK,tr\widetilde{K}\} &\geq &2x,  \label{wave5} \\
\det K+1+L^T\left(
\begin{array}{ll}
y, & z
\end{array}
\right) ^T &\geq &x\cdot trK,  \label{wave6} \\
\det \widetilde{K}+1+\widetilde{L}^T\left(
\begin{array}{ll}
y, & z
\end{array}
\right) ^T &\geq &x\cdot tr\widetilde{K},  \label{wave7}
\end{eqnarray}
where $x=\sqrt{1+y^2+z^2},$ and $L=\left( a-c,2Re(b)\right) $ if we write $K$
as

\begin{equation}
K=\left(
\begin{array}{ll}
a & b \\
b^{*} & c
\end{array}
\right) .
\end{equation}
Ineq.(\ref{wave5}) restricts $(y,z)$ to a circular disk $C$, while Ineq.(\ref
{wave6}) and Ineq.(\ref{wave7}) describe ellipses $E$ and $E^{\prime }$
respectively. The existence of the point $(y,z)$ then turn out to be the
intersection of the ellipses $E$ and $E^{\prime }$ and the circular disk $C$%
. For 3ST state, after lengthy algebra, Ineq.(\ref{wave6}) and Ineq.(\ref
{wave7}) can be written of as
\begin{eqnarray}
t+\frac 1t &\geq &\left( s+\frac 1s\right) x+\frac 13\left( s-\frac
1s\right) y,  \label{wave8} \\
t+\frac 1t &\geq &\left( s+\frac 1s\right) x+\Delta \left( s-\frac 1s\right)
y,  \label{wave9}
\end{eqnarray}
respectively, where we denote $t=2N+1,$ $s=e^{2r}$ and
\begin{equation*}
\Delta =\frac{\left( t+\frac 1t\right) ^2-\frac 43\left[ \left( s+\frac
1s\right) ^2-1\right] }{\left( t+\frac 1t\right) ^2-\frac 49\left[ \left(
s+\frac 1s\right) ^2+5\right] }.
\end{equation*}
Let us consider $\partial E$ and $\partial E^{\prime }$, the borders of $E$ and $%
E^{\prime }$, they are represented by the equality parts of Ineq.(\ref{wave8}%
) and Ineq.(\ref{wave9}).

\begin{eqnarray}
t+\frac 1t &=&\left( s+\frac 1s\right) x+\frac 13\left( s-\frac 1s\right) y,
\label{wave10} \\
t+\frac 1t &=&\left( s+\frac 1s\right) x+\frac \Delta 3\left( s-\frac
1s\right) y.  \label{wave11}
\end{eqnarray}
When $\Delta \neq 1,$ the solution (intersection of $\partial E$ and $%
\partial E^{\prime }$) to above equations will be $y=0$, so that $x=\left(
t+\frac 1t\right) /\left( s+\frac 1s\right) $. While for $\Delta =1,$ we
have $s+\frac 1s=2,$ hence $s=1,$ and we also have $x=\left( t+\frac
1t\right) /\left( s+\frac 1s\right) .$ Because $x\geq 1,$ Ineq.(\ref{wave8})
and Ineq.(\ref{wave9}) then will lead to the condition of fully separable
state, $\left( t+\frac 1t\right) \geq \left( s+\frac 1s\right) ,$ hence $%
t\geq s.$ By denoting $v=\frac N{N+1},$ $\lambda =\tanh r,$ we have a very
simple expression of fully separable condition for 3ST state,
\begin{equation}
v\geq \lambda .  \label{wave12}
\end{equation}
Clearly the two points in $(y,z)$ plane defined by intersection $\partial E$
and $\partial E^{\prime }$ are the farthest points from the origin among all
valid points defined by the intersection $E$ and $E^{\prime }$. To verify
that Ineq. (\ref{wave5}) is also fulfilled, we just need that
\begin{equation}
\min \{trK,tr\widetilde{K}\}\geq 2\frac{t+\frac 1t}{s+\frac 1s}.
\end{equation}
This can be verify numerically for all $v\geq \lambda $, although the
analytical verification is possible. And the conclusion is that Ineq.( )
fulfills for all $v\geq \lambda .$ The districts for fully separable,
biseparable and fully inseparable states are displayed in Fig.(1).

\section{Relative entropy and entanglement bound}

The relative entropy of entanglement for an arbitrary number of parties is
defined by the following formula\cite{Vedral1} \cite{Vedral2}

\begin{equation}
E(\sigma )=\min_{\rho \in D}S(\sigma ||\rho )
\end{equation}
where $D$ is a set of disentangled (separable) states and where $S(\sigma
||\rho )=tr\{\sigma \log \sigma -\sigma \log \rho \}$ is the quantum
relative entropy. For the purpose of this paper, we assume that $D$ is the
set of the states that can be created locally, i.e. it is fully separable
\cite{Vedral2}. Also, by $E_n(\sigma )$ we will always denote the relative
entropy of entanglement for n-party systems with respect to the set of fully
separable states.

Now let us consider the relative entropy of a 3ST state $\sigma $ with
respect to 3ST fully separable state $\rho $. It will be\cite{Chen}
\begin{equation*}
S(\sigma ||\rho )=-S\left( \sigma \right) -3\log \left( 1-v_\rho \right)
+\frac 32\log v_\rho +\frac 12Tr\alpha _\sigma M_\rho .
\end{equation*}
where $S\left( \sigma \right) =-tr\sigma \log \sigma $ is the von Neumann
entropy of $\sigma ,\ $and it can be written of as the $S\left( \sigma
\right) =3g\left( N_\sigma \right) .$ Here $g(x)=\left( x+1\right) \log
(x+10-x\log (x)$ is the\ bosonic entropy function and $N_\sigma =\frac{%
v_\sigma }{1-v_\sigma }$ . By using Eq.(\ref{wave1}), Eq.(\ref{wave2}) and
Eq.(\ref{wave3}) we have

\begin{eqnarray}
S(\sigma ||\rho ) &=&-3g\left( \frac{v_\sigma }{1-v_\sigma }\right) -3\log
\left( 1-v_\rho \right) \\
&&-\frac 32\left[ \frac{1+v_\sigma }{1-v_\sigma }\cosh 2\left( r_\sigma
-r_\rho \right) -1\right] \log v_\rho .  \notag
\end{eqnarray}
For a given 3ST entangled state $\sigma $, the minimum of the relative
entropy $S(\sigma ||\rho )$ will reach at $\lambda _\rho \left( =\tanh
r_\rho \right) =v_\rho $ for some $v_\rho $, that is at the border of fully
separable 3ST state set. The set of fully separable 3ST states is a subset
of $D,$ and we obtain an upper bound $E_{3ur}(\sigma )$ for the relative
entropy of entanglement $E_3(\sigma )$ of 3ST state $\sigma $ as displayed
in Fig.(2).

\begin{eqnarray}
E_{3ur}(\sigma ) &=&\min_{r_\rho }\{-3g\left( \frac{v_\sigma }{1-v_\sigma }%
\right) -3\log \left( 1-\tanh r_\rho \right) \\
&&-\frac 32\left[ \frac{1+v_\sigma }{1-v_\sigma }\cosh 2\left( r_\sigma
-r_\rho \right) -1\right] \log \tanh r_\rho \}.  \notag
\end{eqnarray}
One the other hand, for any pure tripartite state $\sigma $, the
relative entropy of entanglement $E_3(\sigma )$ will also be upper
bounded by teleportation consideration \cite{Vedral2}, that is

\begin{equation}
E_3\left( \sigma \right) \leq \min \{S\left( \sigma _A\right) +S\left(
\sigma _B\right) ,S\left( \sigma _A\right) +S\left( \sigma _C\right)
,S\left( \sigma _B\right) +S\left( \sigma _C\right) \}.
\end{equation}
where $\sigma _A$ is the reduced state of $\sigma $ with $B,C$ parties
traced out. For 3ST pure state $\sigma ,$ it can be simplified to
\begin{equation}
E_3\left( \sigma \right) \leq 2S\left( \sigma _A\right) .
\end{equation}
Hence, one get the upper bound of $E_3\left( \sigma \right) $,
\begin{equation}
E_{3u}\left( \sigma \right) =\min \{E_{3ur}\left( \sigma \right) ,2S\left(
\sigma _A\right) \}\text{, \quad for pure state }\sigma .\text{ }
\end{equation}
The numerical result is displayed in Fig.(3). At high $\lambda \ $side, $%
E_{3ur}\left( \sigma \right) $ is better than $2S\left( \sigma _A\right) $
as upper bound.

\begin{figure}[ptb]
\begin{center}
\includegraphics[
trim=0.000000in 0.000000in -0.138042in 0.000000in,
height=2.0081in, width=2.5097in
]{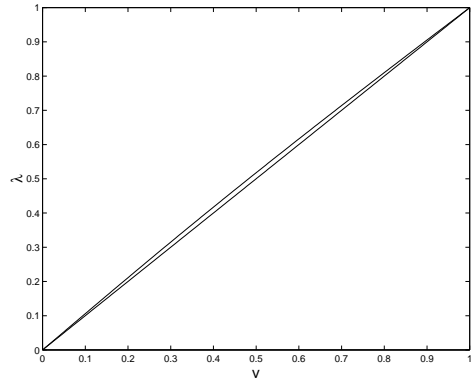}
\end{center}
\caption{The up left district for fully inseparable states, the narrow strip
for biseparable states and the down right triangle district for fully
separable states.}
\end{figure}

\begin{figure}[ptb]
\begin{center}
\includegraphics[
trim=0.000000in 0.000000in -0.138042in 0.000000in,
height=2.0081in, width=2.5097in
]{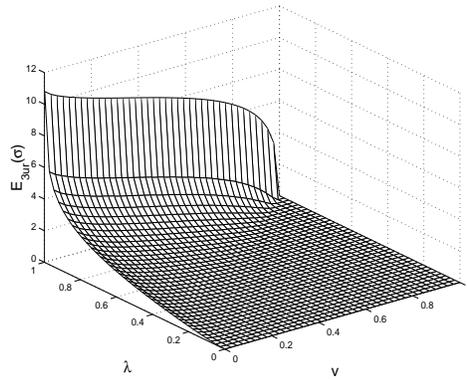}
\end{center}
\caption{Upper bounds of relative entropy of entanglement, with $\protect%
\lambda $ and $v\in [0.001,0.999]$}
\end{figure}

\begin{figure}[ptb]
\begin{center}
\includegraphics[
trim=0.000000in 0.000000in -0.138042in 0.000000in,
height=2.0081in, width=2.5097in
]{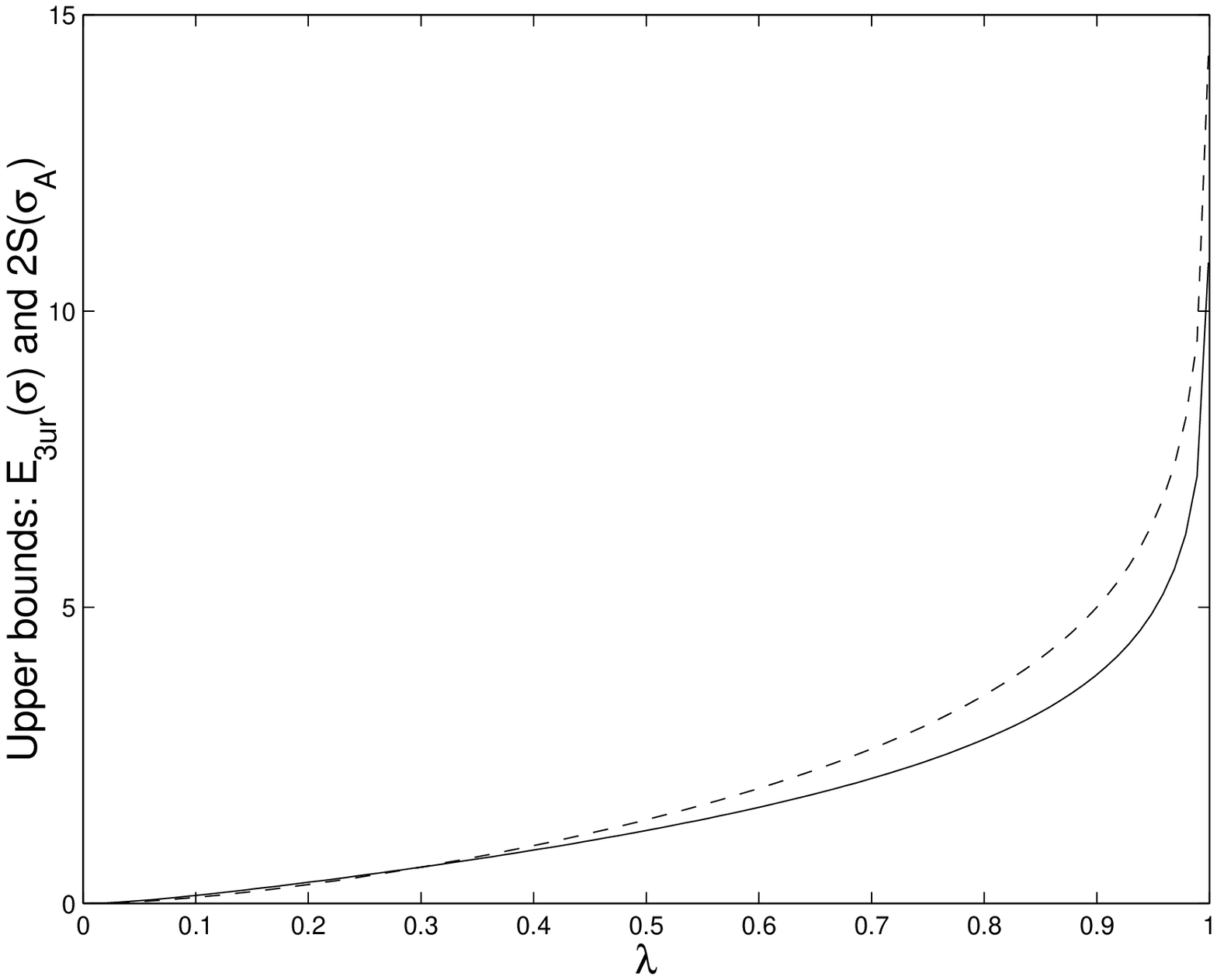}
\end{center}
\caption{Upper bounds of relative entropy of entanglement for pure states
with $\protect\lambda \in [0.001,0.999]$.. The dash line for $2S\left(
\protect\sigma _A\right) ,$ the solid line for $E_{3ur}\left( \protect\sigma
\right) .$}
\end{figure}

\section{Conclusions and Discussions}

The continuous variable analogues of the multipartite entangled
Greenberger-Horne-Zeilinger states had been introduced are pure
states. In this paper we have discussed how to extend these pure
states to the multipartite squeezed thermal states. The such
obtained mST state are highly symmetric and it is characterized by
only two parameters, one is the squeezing parameter and the other
is for thermal noise. I have got the $M$ matrix (matrix in the
exponential expression of density operator of the state) of the
state in order that the relative entropy between the mST states
can be calculated. The criterions of separability have been
simplified for 3ST states. The fully separable condition now takes
a very simple form, that is if the noise is stronger than the
squeezing, the state will be fully separable, otherwise it will be
entangled (biseparable or fully entangled). I have calculated the
relative entropy between the 3ST entangled state and 3ST fully
separable state and obtained an upper bound of the relative
entropy of entanglement for 3ST entangled state.The bound was
compared with the bound obtained by teleportation consideration
for pure 3ST state. Our result is better the the other one at the
high squeezing side.

I in this paper mainly treated 3ST. But there are really some
things need to be talked between the 3ST and 2ST. The bounds
obtained for 3ST are exactly 1.5 times of the bounds for
2ST\cite{ChenQiu}. The separable (fully separable for 3ST
)condition takes exactly the same form. I now wonder if these are
true for mST. I will end this paper with the conjectures for mST:
(1) The fully separable condition will be $v\geq \lambda .$ (2)
The upper bound of relative entropy of entanglement for mST will
be $E_{mur}=\frac m2E_{ur}\left( v,\lambda \right) $ where
$E_{ur}\left( v,\lambda \right) $ is the upper bound of relative
entropy of entanglement for 2ST.

\section{Acknowledgment}

This research was funded by the National Natural Science Foundation of China
(under Grant No. 10347119).

\end{document}